# Unraveling the magnetic softness in Fe-Ni-B based nanocrystalline material by magnetic small-angle neutron scattering

Mathias Bersweiler[a*], Michael P. Adams[a], Inma Peral[a], Joachim Kohlbrecher[b], Kiyonori Suzuki[c], and Andreas Michels[a*]

[a]*Department of Physics and Materials Science, University of Luxembourg, 162A Avenue de la Faïencerie, L-1511 Luxembourg, Grand Duchy of Luxembourg*
[b]*Laboratory for Neutron Scattering, ETH Zurich & Paul Scherrer Institut, 5232 Villigen PSI, Switzerland*
[c]*Department of Materials Science and Engineering, Monash University, Clayton, VIC 3800, Australia*

*Correspondence emails: mathias.bersweiler@uni.lu and andreas.michels@uni.lu*

**Synopsis** We employ magnetic-field-dependent small-angle neutron scattering to analyze the mesoscale magnetic interactions in a soft magnetic HiB-NANOPERM-type alloy and relate the parameters to the experimental coercivity.

**Abstract** We employ magnetic small-angle neutron scattering to investigate the magnetic interactions in $(Fe_{0.7}Ni_{0.3})_{86}B_{14}$ alloy, a HiB-NANOPERM-type soft magnetic nanocrystalline material, which exhibits an ultrafine microstructure with an average grain size below 10 nm. The neutron data reveal a significant spin-misalignment scattering, which is mainly related to the jump of the longitudinal magnetization at internal particle-matrix interfaces. The field dependence of the neutron data can be well described by the micromagnetic small-angle neutron scattering theory. In particular, the theory explains the "clover-leaf-type" angular anisotropy observed in the purely magnetic neutron scattering cross section. The presented neutron-data analysis also provides access to the magnetic interaction parameters, such as the exchange-stiffness constant, which plays a crucial role towards the optimization of the magnetic softness of Fe-based nanocrystalline materials.

## 1. Introduction

Since the pioneering work of Yoshizawa *et al.* (Yoshizawa *et al.*, 1988), the development of novel Fe-based nanocrystalline soft magnetic materials raised considerable interest owing to their great potential for technological applications (Petzold, 2002; Makino *et al.*, 1997). The most well-known examples are FINEMET- (Yoshizawa *et al.*, 1988), VITROPERM- (Vacuumschmelze GmbH, 1993), and NANOPERM-type (Suzuki *et al.*, 1991) soft magnetic alloys, which find widespread application as magnetic cores in high-frequency power transformers or in interface transformers in the ISDN-telecommunication network. For a brief review of the advances in Fe-based nanocrystalline soft magnetic alloys, we refer the reader to the article by Suzuki *et al.* (Suzuki *et al.*, 2019).

More recently, an ultra-fine-grained microstructure combined with excellent soft magnetic properties were obtained in HiB-NANOPERM-type alloys (Li *et al.*, 2020). The magnetic softness in such materials is due to the exchange-averaging effect of the local magnetocrystalline anisotropy $K_1$. This phenomenon has been successfully modeled within the framework of the random anisotropy model (RAM) (Herzer, 1989, 1990; Suzuki *et al.*, 1998; Herzer, 2007), and becomes effective when the average grain size $D$ is smaller than the ferromagnetic exchange length $L_0 = \varphi_0 \sqrt{A_{\mathrm{ex}}/K_1}$, where $A_{\mathrm{ex}}$ is the exchange-stiffness constant and $\varphi_0$ is a proportionality factor of the order of unity which reflects the symmetry of $K_1$. In this regime, the RAM predicts that the coercivity $H_C$ scales as $H_C \propto (D/L_0)^n$, where $n = 3$ or $n = 6$ depending on the nature of the magnetic anisotropy (see, *e.g.*, Refs. (Suzuki *et al.*, 1998, 2019) for details). Therefore, an improvement of the magnetic softness comes about by either reducing $D$ and/or increasing $L_0$.

In the context of increasing $L_0$, the quantitative knowledge of $A_{\mathrm{ex}}$ could help to further develop novel Fe-based soft magnetic nanocrystalline materials. However, up to now, most of the research activities in this field are focused on the overall characterization, *e.g.*, via hysteresis-loop measurements (coercivity, saturation magnetization, and permeability) and magnetic-anisotropy determination (crystalline, shape, or stress related) (McHenry *et al.*, 1999; Herzer, 2013; Suzuki *et al.*, 2019). One reason for this might be related to the fact that many of the conventional methods for measuring $A_{\mathrm{ex}}$ (*e.g.*, magneto-optical, Brillouin light scattering, spin-wave resonance, or inelastic neutron scattering) require thin-film or single-crystal samples.

In the present work, we employ magnetic field-dependent small-angle neutron scattering (SANS) to determine the magnetic interaction parameters in $(Fe_{0.7}Ni_{0.3})_{86}B_{14}$ alloy, specifically, the exchange-stiffness constant and the strength and spatial structure of the magnetic anisotropy and magnetostatic fields. The particular alloy under study is a promising HiB-NANOPERM-type soft magnetic material, which exhibits an ultra-fine microstructure with an average grain size below 10 nm (Li *et al.*, 2020). Magnetic SANS is a unique and powerful technique to investigate the magnetism of materials on the mesoscopic length scale of ∼1–300 nm (*e.g.*, nanorod arrays (Grigoryeva *et al.*, 2007; Günther *et al.*, 2014; Maurer *et al.*, 2014), nanoparticles (Bender *et al.*, 2019; Bersweiler *et al.*, 2019; Zákutná *et al.*, 2020; Kons *et al.*, 2020; Bender *et al.*, 2020; Köhler *et al.*, 2021), INVAR alloy (Stewart *et al.*, 2019) or nanocrystalline materials (Ito *et al.*, 2007; Mettus & Michels, 2015; Titov *et al.*, 2019; Oba *et al.*, 2020; Bersweiler *et al.*, 2021)). For a summary of the fundamentals and the most recent applications of the magnetic SANS technique, we refer the reader to Refs. (Mühlbauer *et al.*, 2019; Michels, 2021).

This paper is organized as follows: Section 2 provides some details of the sample characterization and the neutron experiment. Section 3 summarizes the main expressions for the magnetic SANS cross section and describes the data-analysis procedure to obtain the exchange constant and the average magnetic anisotropy field and magnetostatic field. Section 4 presents and discusses the experimental results, while Section 5 summarizes the main findings of this study.

## 2. Experimental details

Ultra-rapidly annealed $(Fe_{0.7}Ni_{0.3})_{86}B_{14}$ alloy (HiB-NANOPERM-type) was prepared according to the synthesis process detailed in Ref. (Li *et al.*, 2020). The sample for the neutron experiment was prepared by employing the low-capturing isotope $^{11}$B as starting material. The average crystallite size was estimated by wide-angle x-ray diffraction using a Bruker D8 diffractometer in Bragg-Brentano geometry (Cu $K_\alpha$ radiation source). The magnetic measurements were performed at room temperature using a Cryogenic Ltd. vibrating sample magnetometer equipped with a 14 T superconducting magnet and a Riken Denshi BHS-40 DC hysteresis loop tracer. The crystallization and Curie temperatures were determined by means of differential thermal analysis (DTA) and thermo-magneto-gravimetric analysis (TMGA) on Perkin Elmer DTA/TGA 7 analyzers under a constant heating rate of 0.67 K/s. For the neutron experiments, six $(Fe_{0.7}Ni_{0.3})_{86}B_{14}$ ribbons with a surface area of 12 x 20 mm and a thickness of $\approx$ 15 µm were stacked together, resulting in a total sample thickness of $\approx$ 90 µm. The neutron measurements were conducted at the instrument SANS-1 at the Swiss Spallation Neutron Source at the Paul Scherrer Institute, Switzerland. We used an unpolarized incident neutron beam with a mean wavelength of $\lambda$ = 6.0 Å and a wavelength broadening of $\Delta\lambda/\lambda$ = 10% (full width at half maximum). All neutron measurements were conducted at room temperature and within a $q$-range of about 0.036 nm$^{-1}$ $\leq q \leq$ 1.16 nm$^{-1}$. A magnetic field $\boldsymbol{H}_0$ was applied perpendicular to the incident neutron beam ($\boldsymbol{H}_0 \perp \boldsymbol{k}_0$). Neutron data were recorded by decreasing the field from the maximum field available of 8.0 T to 0.02 T following the magnetization curve (see Fig. 2). The internal magnetic field $H_i$ was estimated as $H_i = H_0 - N_d M_S$, where $M_S$ is the saturation magnetization and $N_d$ is the demagnetizing factor, which was determined based on the analytical expression given for a rectangular prism (Aharoni, 1998). The neutron-data reduction (corrections for background scattering and sample transmission) was conducted using the GRASP software package (Dewhurst, 2018).

## 3. Micromagnetic SANS theory

### 3.1. Unpolarized SANS

Based on the micromagnetic SANS theory for two-phase particle-matrix-type ferromagnets developed by Honecker *et al.* (Honecker & Michels, 2013), the elastic total (nuclear + magnetic) unpolarized SANS cross section $d\Sigma/d\Omega$ at momentum-transfer vector $\boldsymbol{q}$ can be formally written as ($\boldsymbol{H}_0 \perp \boldsymbol{k}_0$):

$$\frac{d\Sigma}{d\Omega}(\boldsymbol{q}, H_i) = \frac{d\Sigma_{\mathrm{res}}}{d\Omega}(\boldsymbol{q}) + \frac{d\Sigma_{\mathrm{mag}}}{d\Omega}(\boldsymbol{q}, H_i) \quad (1)$$

where

$$\frac{d\Sigma_{\mathrm{res}}}{d\Omega}(\boldsymbol{q}) = \frac{8\pi^3}{V} b_H^2 \left( b_H^{-2} |\widetilde{N}|^2 + |\widetilde{M}_S|^2 \sin^2(\theta) \right) \quad (2)$$

corresponds to the (nuclear and magnetic) residual SANS cross section, which is measured at complete magnetic saturation, and

$$\begin{aligned} \frac{d\Sigma_{\mathrm{mag}}}{d\Omega}(\boldsymbol{q}, H_i) = \frac{8\pi^3}{V} b_H^2 \Big( & |\widetilde{M}_x|^2 + |\widetilde{M}_y|^2 \cos^2(\theta) \\ & + \left( |\widetilde{M}_z|^2 - |\widetilde{M}_S|^2 \right) \sin^2(\theta) \\ & - \left( \widetilde{M}_y \widetilde{M}_z^* + \widetilde{M}_y^* \widetilde{M}_z \right) \sin(\theta) \cos(\theta) \Big) \end{aligned} \quad (3)$$

denotes the purely magnetic SANS cross-section. In Eqs. (1)–(3), $V$ is the scattering volume, $b_H$ = 2.91 × 10$^8$ A$^{-1}$m$^{-1}$ relates the atomic magnetic moment to the atomic magnetic scattering length, $\widetilde{N}(\boldsymbol{q})$ and $\widetilde{\boldsymbol{M}}(\boldsymbol{q}) = [\widetilde{M}_x(\boldsymbol{q}), \widetilde{M}_y(\boldsymbol{q}), \widetilde{M}_z(\boldsymbol{q})]$ represent the Fourier transforms of the nuclear scattering length density $N(\boldsymbol{r})$ and of the magnetization vector field $\boldsymbol{M}(\boldsymbol{r})$, respectively, $\theta$ specifies the angle between $\boldsymbol{H}_0$ and $\boldsymbol{q} \cong q\{0, \sin(\theta), \cos(\theta)\}$ in the small-angle approximation, and the asterisks "*" denote the complex conjugated quantities. $\widetilde{M}_S(\boldsymbol{q})$ is the Fourier transform of the saturation magnetization profile $M_S(\boldsymbol{r})$, *i.e.*, $\widetilde{M}_S(\boldsymbol{q}) = \widetilde{M}_z(\boldsymbol{q})$ at complete magnetic saturation [compare Eq. (2)]. For small-angle scattering, the component of the scattering vector along the incident neutron beam, here $q_x$, is smaller than the other two components $q_y$ and $q_z$, so that only correlations in the plane perpendicular to the incoming neutron beam are probed.

In our neutron-data analysis, to experimentally access the $d\Sigma_{\mathrm{mag}}/d\Omega$, we have subtracted the SANS cross section $d\Sigma/d\Omega$ measured at the largest available field (approach-to-saturation regime; compare Fig. 2) from the $d\Sigma/d\Omega$ measured at lower fields. This specific subtraction procedure eliminates the nuclear SANS contribution $\propto |\widetilde{N}|^2$, which is field independent, and therefore

$$\frac{d\Sigma_{\mathrm{mag}}}{d\Omega}(\boldsymbol{q},H_i) = \frac{8\pi^3}{V} b_H^2 \Big(\Delta\big|\widetilde{M}_x\big|^2 + \Delta\big|\widetilde{M}_y\big|^2 \cos^2(\theta) + \Delta\big|\widetilde{M}_z\big|^2 \sin^2(\theta) - \Delta\big(\widetilde{M}_y \widetilde{M}_z^* + \widetilde{M}_y^* \widetilde{M}_z\big)\sin(\theta)\cos(\theta)\Big), \quad (4)$$

where the "Δ" represent the differences of the Fourier components at the two selected fields (low field minus highest field).

### 3.2. Approach-to-saturation regime

In the particular case of the approach-to-saturation regime, where $\widetilde{M}_z \cong \widetilde{M}_S$, and which implies therefore $\Delta\big|\widetilde{M}_z\big|^2 \to 0$ in Eq. (4), $d\Sigma/d\Omega$ can be re-written as:

$$\frac{d\Sigma}{d\Omega}(\boldsymbol{q},H_i) = \frac{d\Sigma_{\mathrm{res}}}{d\Omega}(\boldsymbol{q}) + S_H(\boldsymbol{q}) \times R_H(\boldsymbol{q},H_i) + S_M(\boldsymbol{q}) \times R_M(\boldsymbol{q},H_i) \quad (5)$$

where $S_H(\boldsymbol{q}) \times R_H(\boldsymbol{q},H_i)$ and $S_M(\boldsymbol{q}) \times R_M(\boldsymbol{q},H_i)$ correspond to the magnetic scattering contributions due to perturbing magnetic anisotropy fields and magnetostatic fields, respectively. More specifically, the anisotropy-field scattering function

$$S_H(\boldsymbol{q}) = \frac{8\pi^3}{V} b_H^2 \big|\widetilde{H}_p(\boldsymbol{q})\big|^2 \quad (6)$$

depends of the Fourier coefficient $\widetilde{\boldsymbol{H}}_p(\boldsymbol{q})$ of the magnetic anisotropy field, whereas the scattering function of the longitudinal magnetization

$$S_M(\boldsymbol{q}) = \frac{8\pi^3}{V} b_H^2 \big|\widetilde{M}_z(\boldsymbol{q})\big|^2 \quad (7)$$

is related to the Fourier coefficient $\widetilde{M}_z \propto \Delta M$. For an inhomogeneous material of the NANOPERM-type, the latter quantity is related to the magnetization jump $\Delta M$ at internal (*e.g.*, particle-matrix) interfaces. We would like to emphasize that the $\boldsymbol{q}$ dependence of $S_H$ and $S_M$ can often be described by a particle form factor (*e.g.*, sphere) or a Lorentzian-squared function. The corresponding (dimensionless) micromagnetic response functions $R_H$ and $R_M$ are given by:

$$R_H(\boldsymbol{q},H_i) = \frac{p^2}{2}\left[1 + \frac{\cos^2\theta}{(1 + p\sin^2\theta)^2}\right] \quad (8)$$

and

$$R_M(\boldsymbol{q},H_i) = \frac{p^2 \sin^2\theta \cos^4\theta}{(1 + p\sin^2\theta)^2} + \frac{2p\sin^2\theta\cos^2\theta}{1 + p\sin^2\theta}. \quad (9)$$

The dimensionless function $p(q,H_i) = M_S/[H_i(1 + l_H^2 q^2)]$ depends on the internal magnetic field $H_i$ and on the exchange length $l_H(H_i) = \sqrt{2A_{\mathrm{ex}}/(\mu_0 M_S H_i)}$.

### 3.3. Estimation of the magnetic interaction parameters

Most of the time it is more convenient to analyze the (over 2π) azimuthally-averaged SANS cross sections instead of the two-dimensional ones. By performing an azimuthal average of the response functions [Eqs. (8) and (9)] with respect to the angle $\theta$, *i.e.*, $1/(2\pi)\int_0^{2\pi}(\dots)d\theta$ and by assuming $S_H$ and $S_M$ to be isotropic ($\theta$-independent), the SANS cross section $d\Sigma/d\Omega$ can be written as:

$$\frac{d\Sigma}{d\Omega}(q,H_i) = \frac{d\Sigma_{res}}{d\Omega}(q) + S_H(q) \times R_H(q,H_i) + S_M(q) \times R_M(q,H_i) \quad (10)$$

where

$$R_H(q,H_i) = \frac{p^2}{4}\left[2 + \frac{1}{\sqrt{1+p}}\right] \quad (11)$$

and

$$R_M(q,H_i) = \frac{\sqrt{1+p} - 1}{2}. \quad (12)$$

For a given set of parameters $A_{\mathrm{ex}}$ and $M_S$, the numerical values of $R_H$ and $R_M$ are known at each value of $q$ and $H_i$. Because of the linearity of Eq. (10) in $R_H$ and $R_M$, one can obtain the values of $d\Sigma_{\mathrm{res}}/d\Omega$ (as the intercept) and $S_H$ and $S_M$ (as the slopes) at each $q$-value by performing a (weighted) non-negative least-squares global fit of the azimuthally-averaged SANS cross sections $d\Sigma/d\Omega$ measured at several $H_i$. By treating $A_{\mathrm{ex}}$ in the expression for $p(q,H_i)$ as an adjustable parameter during the fitting procedure allows us to estimate this quantity. The best-fit value for $A_{\mathrm{ex}}$ is obtained from the minimization of the (weighted) mean-squared deviation between experiment and fit:

$$\chi^2(A_{\mathrm{ex}}) = \frac{1}{N}\sum_{\mu=1}^{N_\mu}\sum_{\nu=1}^{N_\nu}\frac{1}{\sigma_{\mu,\nu}^2}\left[\frac{d\Sigma^{\mathrm{exp}}}{d\Omega}(q_\mu,H_{i,\nu}) - \frac{d\Sigma^{\mathrm{sim}}}{d\Omega}(q_\mu,H_{i,\nu})\right]^2 \quad (13)$$

where the indices $\mu$ and $\nu$ refer to the particular $q$ and $H_i$-values, the $\sigma_{\mu,\nu}^2$ denote the uncertainties in the experimental data, $N = N_\mu N_\nu$ corresponds to the number of data points, and $d\Sigma^{\mathrm{exp}}/d\Omega$ and $d\Sigma^{\mathrm{sim}}/d\Omega$ are the azimuthally-averaged SANS cross section determined from the neutron experiments and numerically computed using Eq. (10), respectively. We would like to point out that the best-fit value for $A_{\mathrm{ex}}$ represents an average over the sample volume.

Finally, the numerical integration of the determined $S_H(q)$ and $S_M(q)$ over the whole-$\boldsymbol{q}$ space according to (Honecker & Michels, 2013)

$$\frac{1}{2\pi^2 b_H^2}\int_0^\infty S_{H,M}(q) q^2 dq \quad (14)$$

yields the mean-square anisotropy field $\langle |H_p|^2 \rangle$ and the mean-square longitudinal magnetization fluctuation $\langle |M_z|^2 \rangle$, respectively. Since the neutron experiments are performed within a finite $q$-range and since both integrands $S_{H,M} q^2$ do not exhibit any sign of convergence, one can only obtain a lower bound for both quantities by numerical integration. Moreover, it is important to realize that the specific neutron-data analysis described above does not represent a "continuous" fit of $d\Sigma/d\Omega$ in the conventional sense, but rather the point by point reconstruction of the theoretical cross sections based on the experimental data.

## 4. Results and discussion

Figure 1 displays the wide-angle x-ray diffraction (XRD) results of the $(Fe_{0.7}Ni_{0.3})_{86}B_{14}$ ribbons. The XRD pattern exhibits only the reflections from the fcc-Fe(Ni) phase, as expected for this particular composition (Li *et al.*, 2020), and therefore confirms the high-quality synthesis of the sample. The values of the lattice parameter $a$ and the average crystallite size $D$ were estimated from the XRD data refinement using the LeBail fit method (LBF) implemented in the Fullprof Suite (Rodríguez-Carvajal, 1993). The best-fit values are summarized in Table 1. Both values are consistent with data in the literature (compare Refs. (Anand *et al.*, 2019) and Ref. (Li *et al.*, 2020) for $a$ and $D$, respectively). As previously discussed, the origin of the exceptionally fine microstructure observed in $(Fe_{0.7}Ni_{0.3})_{86}B_{14}$ alloys may be qualitatively attributed to the ultrafast nucleation kinetics of the fcc-Fe(Ni) phase (Li *et al.*, 2020).

Figure 2(a) presents the positive magnetization branch on a semi-logarithmic scale (measured at room temperature), while the hysteresis loop on a linear-linear scale, and between $\pm$ 0.03 mT, is displayed in Fig. 2(b). The data have been normalized by the saturation magnetization $M_S$, which was estimated from the linear regression $M(1/H_i)$ for $\mu_0 H_i \in [10\,\mathrm{T} - 14\,\mathrm{T}]$ [see inset in Fig. 2(a)]. The values of $M_S$ and $H_C$ (see Table 1) are in agreement with the ones reported in the literature (Li *et al.*, 2020). Defining the approach-to-saturation regime by $M/M_S \geq 90$ %, we can see that this regime is reached for $\mu_0 H_i \geq 65$ mT. Moreover, the extremely small value for $H_C$ combined with the high $M_S$ confirms the huge potential of $(Fe_{0.7}Ni_{0.3})_{86}B_{14}$ alloy as a soft magnetic material, and suggests that in the framework of the RAM (Herzer, 2007), $H_C$ should fall into the regime where $H_C \propto (D/L_0)^3$ (Suzuki *et al.*, 2019).

Figure 3 shows the DTA and TMGA curves for amorphous $(Fe_{0.7}Ni_{0.3})_{86}B_{14}$ alloy. Two exothermic peaks are evident on the DTA curve reflecting the well-known two-stage reactions, where fcc-Fe(Ni) forms at the 1st peak followed by decomposition of the residual amorphous phase at the 2nd peak. The sharp drop of the TMGA signal just before the 2nd stage crystallization corresponds to the Curie temperature of the residual amorphous phase ($T_C^{\mathrm{am}} \approx 720$ K). This value, which reflects the exchange integral in our sample (see below), is consistent with the ones determined for amorphous $Fe_{86}B_{14}$ samples prepared under similar conditions (Zang *et al.*, 2020).

Figure 4 (upper row) shows the experimental two-dimensional total (nuclear + magnetic) SANS cross sections $d\Sigma/d\Omega$ of the $(Fe_{0.7}Ni_{0.3})_{86}B_{14}$ ribbons at different selected fields. As can be seen, at $\mu_0 H_i = 7.99$ T (near saturation), the pattern is predominantly elongated perpendicular to the magnetic field direction. This particular feature in $d\Sigma/d\Omega$ is the signature of the so-called "$\sin^2(\theta)$-type" angular anisotropy [compare Eq. (2)]. Near saturation, the magnetic scattering resulting from the spin misalignment is small compared to the one resulting from the longitudinal magnetization jump at the internal (*e.g.*, particle-matrix) interfaces. By reducing the field, the patterns remain predominantly elongated perpendicular to the magnetic field, but at the smaller momentum transfers $q$ an additional field-dependent signal is observed "roughly" along the diagonals of the detector, suggesting a more complex magnetization structure. Figure 4 (middle row) presents the corresponding two-dimensional purely magnetic SANS cross sections $d\Sigma_{\mathrm{mag}}/d\Omega$ determined by subtracting $d\Sigma/d\Omega$ at $\mu_0 H_i = 7.99$ T from the data at lower fields. In this way, the maxima along the diagonals of the detector become more clearly visible, thereby revealing the so-called "clover-leaf-type" angular anisotropy pattern. This particular feature was also previously observed in NANOPERM-type soft magnetic materials (Honecker *et al.*, 2013), and is related to the dominant magnetostatic term $S_M \times R_M$ in the expression for $d\Sigma_{\mathrm{mag}}/d\Omega$ [compare Eqs. (8) and (9)]. More specifically, the jump in the magnitude of the saturation magnetization at the particle-matrix interfaces, which can be of the order of 1 T in these type of alloys (Honecker *et al.*, 2013), results in dipolar stray fields which produce spin disorder in the surroundings. Figure 4 (lower row) displays $d\Sigma_{\mathrm{mag}}/d\Omega$ computed using the micromagnetic SANS theory [Eqs. (5)–(9)] and the experimental parameters summarized in Table 1. As is seen, the clover-leaf-type angular anisotropy experimentally observed in Fig. 4 (middle row) can be well reproduced using the micromagnetic theory.

Figure 5(a) displays the (over 2π) azimuthally-averaged $d\Sigma/d\Omega$, while the corresponding $d\Sigma_{\mathrm{mag}}/d\Omega$ are shown in Fig. 5(b). By decreasing $\mu_0 H_i$ from 7.99 T to 10 mT, the intensity of $d\Sigma/d\Omega$ increases by almost two orders of magnitude at the smallest momentum transfers $q$. By comparison to Eqs. (1)–(4), it appears obvious that the magnetic-field dependence of $d\Sigma/d\Omega$ can only result from the mesoscale spin disorder (*i.e.*, from the failure of the spins to be fully aligned along $\boldsymbol{H}_0$). As is seen in Fig. 5(b), the magnitude of $d\Sigma_{\mathrm{mag}}/d\Omega$ is of the same order as $d\Sigma/d\Omega$, supporting therefore the notion of dominant spin-misalignment scattering in $(Fe_{0.7}Ni_{0.3})_{86}B_{14}$ alloy.

Figure 6 shows the magnetic SANS results determined from the field-dependent approach described in Sec. 3.3. In the present case, to warrant the validity of the micromagnetic SANS theory, only $d\Sigma/d\Omega$ measured for $\mu_0 H_i \geq 65$ mT were considered (*i.e.*, within the approach-to-saturation regime, compare Fig. 2). We have also restricted our neutron-data analysis to $q \leq q_{\max} = \sqrt{\mu_0 M_S H_0^{\max}/(2A_{\mathrm{ex}})} = 0.65$ nm$^{-1}$, since the magnetic SANS cross section is expected to be field-independent for $q \geq q_{\max}$ (Michels, 2021). In Fig. 6(a), we plot the (over 2π) azimuthally-averaged $d\Sigma/d\Omega$ along with the corresponding fits based on the micromagnetic SANS theory [Eq. (10), black solid lines]. It is seen that the field dependence of $d\Sigma/d\Omega$ over the restricted $q$-range can be well reproduced by the theory. Figure 6(b) displays the (weighted) mean-squared deviation between experiment and fit, $\chi^2$, determined according to Eq. (13), as a function of the exchange-stiffness constant $A_{ex}$. In this way, we find $A_{\mathrm{ex}} = (10 \pm 1)$ pJ/m (see Table 1). The comparison with previous studies is discussed in the next paragraph for more clarity. Figure 6(c) displays the best-fit results for $d\Sigma_{\mathrm{res}}/d\Omega$, $S_H$, and $S_M$. Not surprisingly, the magnitude of $d\Sigma_{\mathrm{res}}/d\Omega$ (limit of $d\Sigma/d\Omega$ at infinite field) is smaller than the $d\Sigma/d\Omega$ at the largest fields [compare Fig. 6(a)], supporting the validity of the micromagnetic SANS theory. Furthermore, the magnitude of $S_H$ is about two orders of magnitude smaller than $S_M$, suggesting that the magnetization jump $\Delta M$ at internal particle-matrix interfaces represents the main source of spin disorder in this material. The estimated values for the mean-square anisotropy field and the mean-square magnetostatic field in terms of Eq. (14) are, respectively, 0.3 mT and 24 mT. These values qualitatively support the notion of dominant spin-misalignment scattering due to magnetostatic fluctuations. The $q$-dependence of $S_M$ can be described using a Lorentzian-squared function [blue solid line in Fig. 6(c)] from which an estimate for the magnetostatic correlation length $\xi_M = 2.4 \pm 0.2$ nm is obtained. This value compares favorably with the value of $l_M = \sqrt{2A_{\mathrm{ex}}/\mu_0 M_S^2} = 3.7$ nm (using $A_{\mathrm{ex}} = 10$ pJ/m and $\mu_0 M_S = 1.34$ T [taken from Table 1]), which reflects the competition between the exchange and magnetostatic energies.

We would like to emphasize that our experimental value for $A_{\mathrm{ex}} = 10$ pJ/m is about 2–3 times larger than the ones reported in NANOPERM-type soft magnetic materials (Honecker *et al.*, 2013). Since the Curie temperature of the residual amorphous phase in our nanocrystalline $(Fe_{0.7}Ni_{0.3})_{86}B_{14}$ sample is well above 700 K (see Fig. 3 and Table 1), while the one in the $Fe_{89}Zr_7B_3Cu_1$ sample used in the previous study (Honecker *et al.*, 2013) was as low as 350 K, the local exchange stiffness in the grain boundary amorphous phase in HiB-NANOPERM-type alloys is expected to be higher than the one in NANOPERM-type alloys. This finding could explain the origin of the larger $A_{\mathrm{ex}}$ value reported in the present study. Therefore, one can expect an improvement of the magnetic softness in HiB-NANOPERM thanks to the ensuing increase of the ferromagnetic exchange length $L_0$. It is well established that nonmagnetic and/or ferromagnetic additives and the annealing conditions strongly affect the microstructural and magnetic properties of Fe-based nanocrystalline materials (McHenry *et al.*, 1999; Herzer, 2007, 2013; Suzuki *et al.*, 2019) and therefore have a strong impact on their magnetic softness. Using $A_{\mathrm{ex}} = 10$ pJ/m (this study), $K_1 \approx 9.0$ kJ/m$^3$ [footnote 1], and $\varphi_0 \approx 1.5$ (Herzer, 2007), we obtain $L_0 \approx 50$ nm. This value for $L_0$ is in very good agreement with the typical length scale of $\sim 30 - 50$ nm previously reported in soft magnetic Fe-based alloys. Moreover, the comparison of the average grain size $D = 7$ nm with the $L_0$ value, here $D \ll L_0$, also confirms that in the framework of the random anisotropy model (Herzer, 1989, 1990; Suzuki *et al.*, 1998; Herzer, 2007), the exchange-averaged magnetic anisotropy $\langle K \rangle$ falls into the regime where $\langle K \rangle \propto D^3$. This finding is also consistent with the (experimental) $D^3$-dependence of $H_C$ reported in Fe-B-based HiB-NANOPERM alloys (Suzuki *et al.*, 2019; Li *et al.*, 2020).

## 5. Conclusions

We employed magnetic SANS to determine the magnetic interaction parameters in $(Fe_{0.7}Ni_{0.3})_{86}B_{14}$ alloy, which is a HiB-NANOPERM-type soft magnetic material. The analysis of the magnetic SANS data suggests the presence of strong spin misalignment on a mesoscopic length scale. In fact, the micromagnetic SANS theory provides an excellent description of the field dependence of the total (nuclear + magnetic) and purely magnetic SANS cross sections. The clover-leaf-type angular anisotropy patterns observed in the magnetic SANS signal can be well reproduced by the theory. The magnitudes of the scattering functions $S_H$ and $S_M$ allow one to conclude that the magnetization jumps at internal particle-matrix interfaces and the ensuing dipolar stray fields are the main source of the spin-disorder in this material. Our study highlights the strength of the magnetic SANS technique to characterize magnetic materials on the mesoscopic length scale. The structural and magnetic results (summarized in Table 1) provide valuable information on the $(Fe_{0.7}Ni_{0.3})_{86}B_{14}$ ribbons, and further confirm the strong potential of Fe-Ni-B-based HiB-NANOPERM-type alloys as soft magnetic nanocrystalline materials. In the context of the random anisotropy model, we demonstrated that the magnetic softness in this system is due to the combined action of the small particle size ($D = 7$ nm) and an increased exchange constant ($A_{\mathrm{ex}} = 10$ pJ/m) resulting in an enhanced exchange correlation length $L_0$.

The data that support the findings of this study are available from the corresponding author upon reasonable request.

## Acknowledgements

The authors acknowledge the Swiss spallation neutron source at the Paul Scherrer Institute, Switzerland, for the provision of neutron beamtime. A.M. and M.B. acknowledge financial support from the National Research Fund of Luxembourg.

## References

Aharoni, A. (1998). *J. Appl. Phys.* **83**, 3432–3434.

Anand, K. S., Goswami, D., Jana, P. P. & Das, J. (2019). *AIP Adv.* **9**, 055126.

Bender, P., Honecker, D. & Fernández Barquín, L. (2019). *Appl. Phys. Lett.* **115**, 132406.

Bender, P., Marcano, L., Orue, I., Venero, D. A., Honecker, D., Fernández Barquín, L., Muela, A. & Luisa Fdez-Gubieda, M. (2020). *Nanoscale Adv.* **2**, 1115–1121.

Bersweiler, M., Bender, P., Vivas, L. G., Albino, M., Petrecca, M., Mühlbauer, S., Erokhin, S., Berkov, D., Sangregorio, C. & Michels, A. (2019). *Phys. Rev. B.* **100**, 144434.

Bersweiler, M., Sinaga, E. P., Peral, I., Adachi, N., Bender, P., Steinke, N. J., Gilbert, E. P., Todaka, Y., Michels, A. & Oba, Y. (2021). *Phys. Rev. Mat.* **5**, 044409.

Dewhurst, C. D. (2018). Graphical Reduction and Analysis SANS Program for Matlab™, https://www.ill.eu/users/support-labs-infrastructure/software-scientific-tools/grasp (Institut Laue–Langevin, Grenoble, 2018).

Grigoryeva, N. A., Grigoriev, S. V., Eckerlebe, H., Eliseev, A. A., Lukashin, A. V. & Napolskii, K. S. (2007). *J. Appl. Crystallogr.* **40**, s532–s536.

Günther, A., Bick, J.-P., Szary, P., Honecker, D., Dewhurst, C. D., Keiderling, U., Feoktystov, A. V., Tschöpe, A., Birringer, R. & Michels, A. (2014). *J. Appl. Crystallogr.* **47**, 992–998.

Hall, R. C. (1960). *J. Appl. Phys.* **31**, 1037–1038.

Herzer, G. (1989). *IEEE Trans. Magn.* **25**, 3327–3329.

Herzer, G. (1990). *IEEE Trans. Magn.* **26**, 1397–1402.

Herzer, G. (2007). *Handbook of Magnetism and Advanced Magnetic Materials*, Vol. *4*, edited by H. Kronmüller & S. Parkin, pp. 1882–1908. John Wiley & Sons, Ltd.

Herzer, G. (2013). *Acta Mater.* **61**, 718–734.

Honecker, D., Dewhurst, C. D., Suzuki, K., Erokhin, S. & Michels, A. (2013). *Phys. Rev. B*. **88**, 094428.

Honecker, D. & Michels, A. (2013). *Phys. Rev. B*. **87**, 224426.

Ito, N., Michels, A., Kohlbrecher, J., Garitaonandia, J. S., Suzuki, K. & Cashion, J. D. (2007). *J. Magn. Magn. Mater.* **316**, 458–461.

Köhler, T., Feoktystov, A., Petracic, O., Kentzinger, E., Bhatnagar-Schöffmann, T., Feygenson, M., Nandakumaran, N., Landers, J., Wende, H., Cervellino, A., Rücker, U., Kovács, A., Dunin-Borkowski, R. E. & Brückel, T. (2021). *Nanoscale*. **13**, 6965–6976.

Kons, C., Phan, M. H., Srikanth, H., Arena, D. A., Nemati, Z., Borchers, J. A. & Krycka, K. L. (2020). *Phys. Rev. Mater.* **4**, 034408.

Li, Z., Parsons, R., Zang, B., Kishimoto, H., Shoji, T., Kato, A., Karel, J. & Suzuki, K. (2020). *Scr. Mater.* **181**, 82–85.

Makino, A., Hatanai, T., Naitoh, Y., Bitoh, T., Inoue, A. & Masumoto, T. (1997). *IEEE Trans. Magn.* **33**, 3793–3797.

Maurer, T., Gautrot, S., Ott, F., Chaboussant, G., Zighem, F., Cagnon, L. & Fruchart, O. (2014). *Phys. Rev. B*. **89**, 184423.

McHenry, M. E., Willard, M. A. & Laughlin, D. E. (1999). *Prog. Mater. Sci.* **44**, 291–433.

Mettus, D. & Michels, A. (2015). *J. Appl. Crystallogr.* **48**, 1437–1450.

Michels, A. (2021). Magnetic Small-Angle Neutron Scattering: A Probe for Mesoscale magnetism Analysis Oxford: Oxford University Press.

Mühlbauer, S., Honecker, D., Périgo, E. A., Bergner, F., Disch, S., Heinemann, A., Erokhin, S., Berkov, D., Leighton, C., Eskildsen, M. R. & Michels, A. (2019). *Rev. Mod. Phys.* **91**, 015004.

Oba, Y., Adachi, N., Todaka, Y., Gilbert, E. P. & Mamiya, H. (2020). *Phys. Rev. Res.* **2**, 033473.

Petzold, J. (2002). *J. Magn. Magn. Mater.* **242**–**245**, 84–89.

Rodríguez-Carvajal, J. (1993). *Phys. B*. **192**, 55–69.

Stewart, J. R., Giblin, S. R., Honecker, D., Fouquet, P., Prabhakaran, D. & Taylor, J. W. (2019). *J. Phys. Condens. Matter*. **31**, 025802.

Suzuki, K., Herzer, G. & Cadogan, J. M. (1998). *J. Magn. Magn. Mater.* **177**–**181**, 949–950.

Suzuki, K., Makino, A., Inoue, A. & Masumoto, T. (1991). *J. Appl. Phys.* **70**, 6232–6237.

Suzuki, K., Parsons, R., Zang, B., Onodera, K., Kishimoto, H., Shoji, T. & Kato, A. (2019). *AIP Adv.* **9**, 035311.

Tarasov, L. P. (1939). *Phys. Rev.* **56**, 1245.

Titov, I., Barbieri, M., Bender, P., Peral, I., Kohlbrecher, J., Saito, K., Pipich, V., Yano, M. & Michels, A. (2019). *Phys. Rev. Mater.* **3**, 84410.

Vacuumschmelze GmbH (1993). *Toroidal Cores of VITROPERM,* data sheet PW-014.

Yoshizawa, Y., Oguma, S. & Yamauchi, K. (1988). *J. Appl. Phys.* **64**, 6044–6046.

Zákutná, D., Nižňanský, D., Barnsley, L. C., Babcock, E., Salhi, Z., Feoktystov, A., Honecker, D. & Disch, S. (2020). *Phys. Rev. X*. **10**, 031019.

Zang, B., Parsons, R., Onodera, K., Kishimoto, H., Shoji, T., Kato, A., Garitaonandia, J. S., Liu, A. C. Y. & Suzuki, K. (2020). *Phys. Rev. Mater.* **4**, 33404.

# FIGURE 1

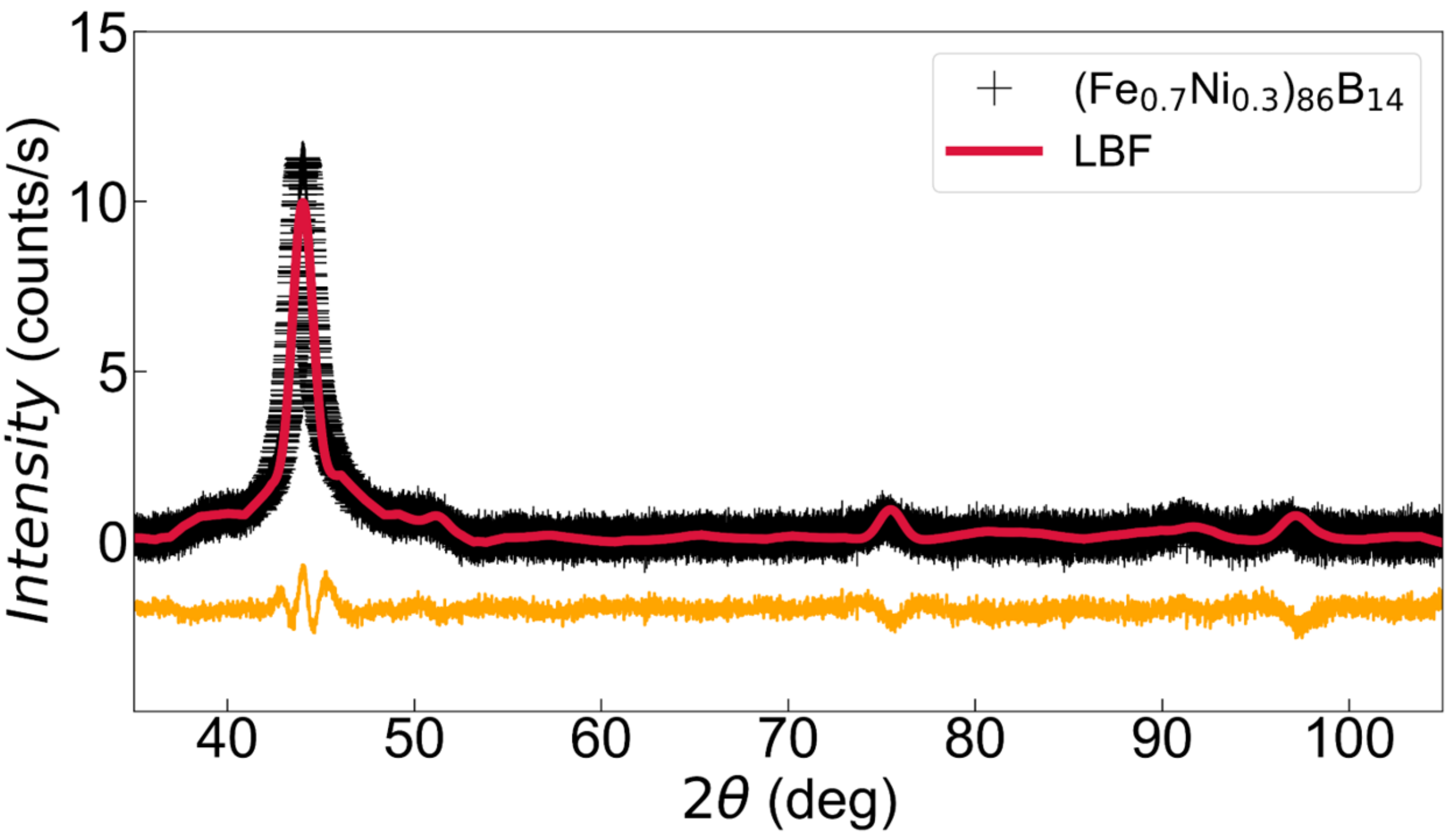

Figure 1: X-ray diffraction (XRD) pattern for $(Fe_{0.7}Ni_{0.3})_{86}B_{14}$ ribbons, a HiB-NANOPERM-type soft magnetic nanocrystalline material (black crosses; Cu $K_{\alpha}$ radiation). Red solid line: XRD data refinement using the Le Bail fit method (LBF) implemented in the Fullprof software. The bottom orange solid line represents the difference between the calculated and experimental intensities.

# FIGURE 2

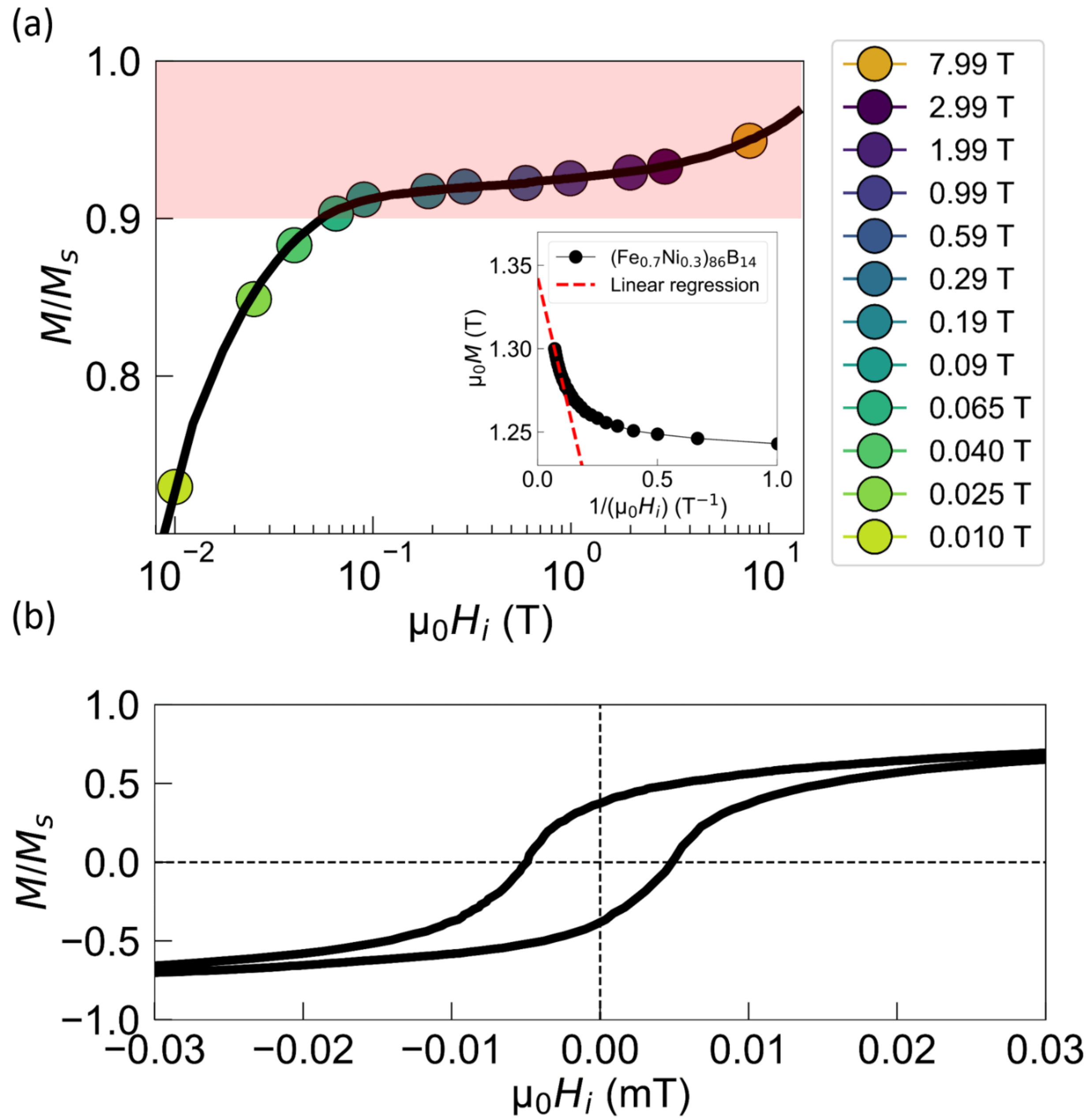

Figure 2: (a) Normalized positive magnetization branch measured at room temperature (semi-logarithmic scale). Color-filled circles: $M/M_s$ values for which the small-angle neutron scattering measurements have been performed. The approach-to saturation regime, defined as $M/M_s \geq 90\%$, is indicated by the red-shaded area. Inset: Plot of the magnetization as a function of $1/H_i$ (black circles). Red dashed line: Linear regression for $\mu_0 H_i \in [10\ \mathrm{T} - 14\ \mathrm{T}]$ (linear – linear scale). (b) Normalized magnetization curve measured using a Riken Denshi BHS-40 DC hysteresis loop tracer, revealing a coercivity of $\mu_0 H_C \approx 0.0049$ mT (linear-linear scale).

# FIGURE 3

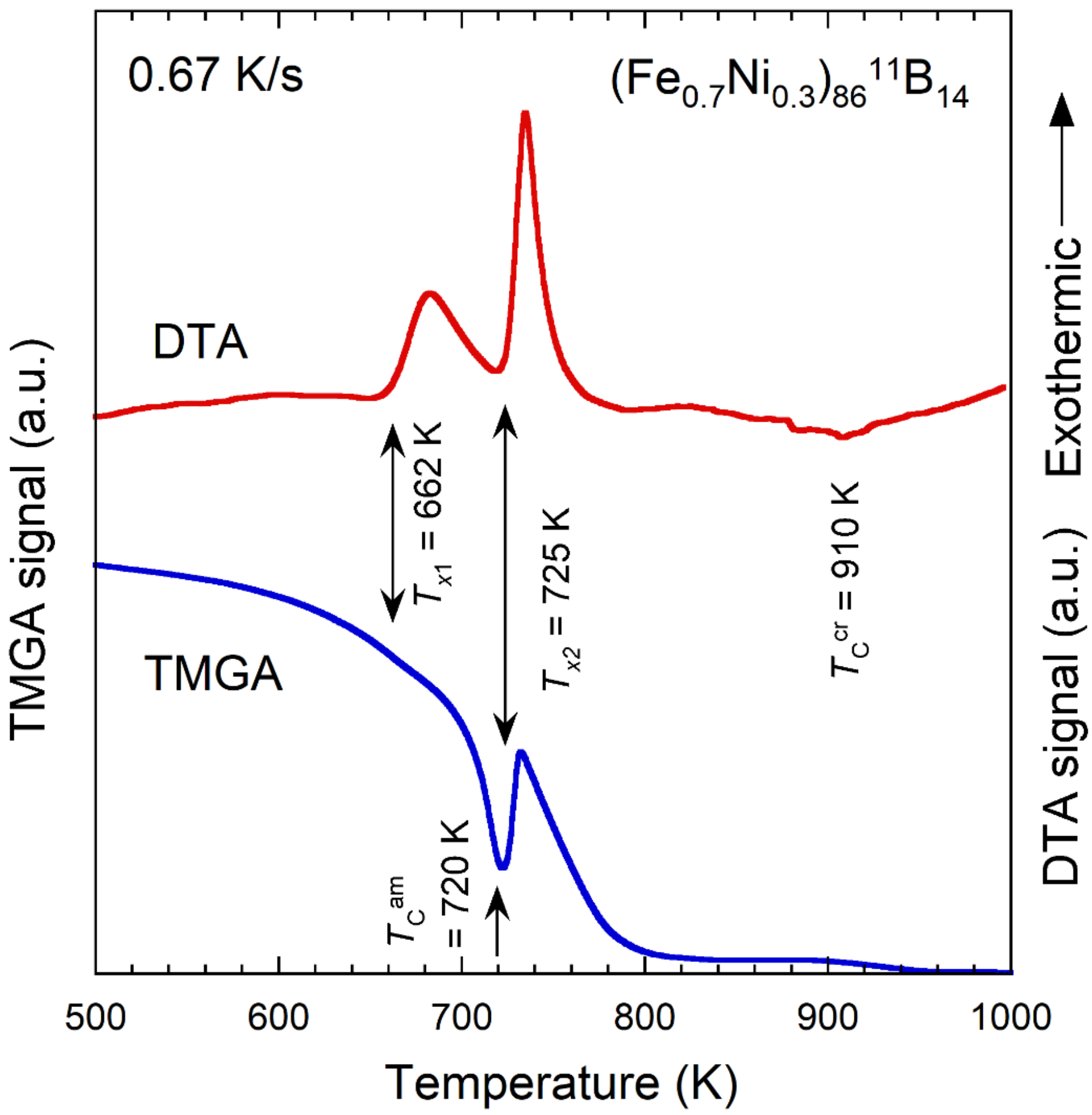

Figure 3: Results of differential thermal analysis (DTA) (red solid line) and thermo-magneto-gravimetric analysis (TMGA) (blue solid line) for amorphous $(Fe_{0.7}Ni_{0.3})_{86}B_{14}$ alloy. The arrows mark the crystallization and Curie temperatures.

# FIGURE 4

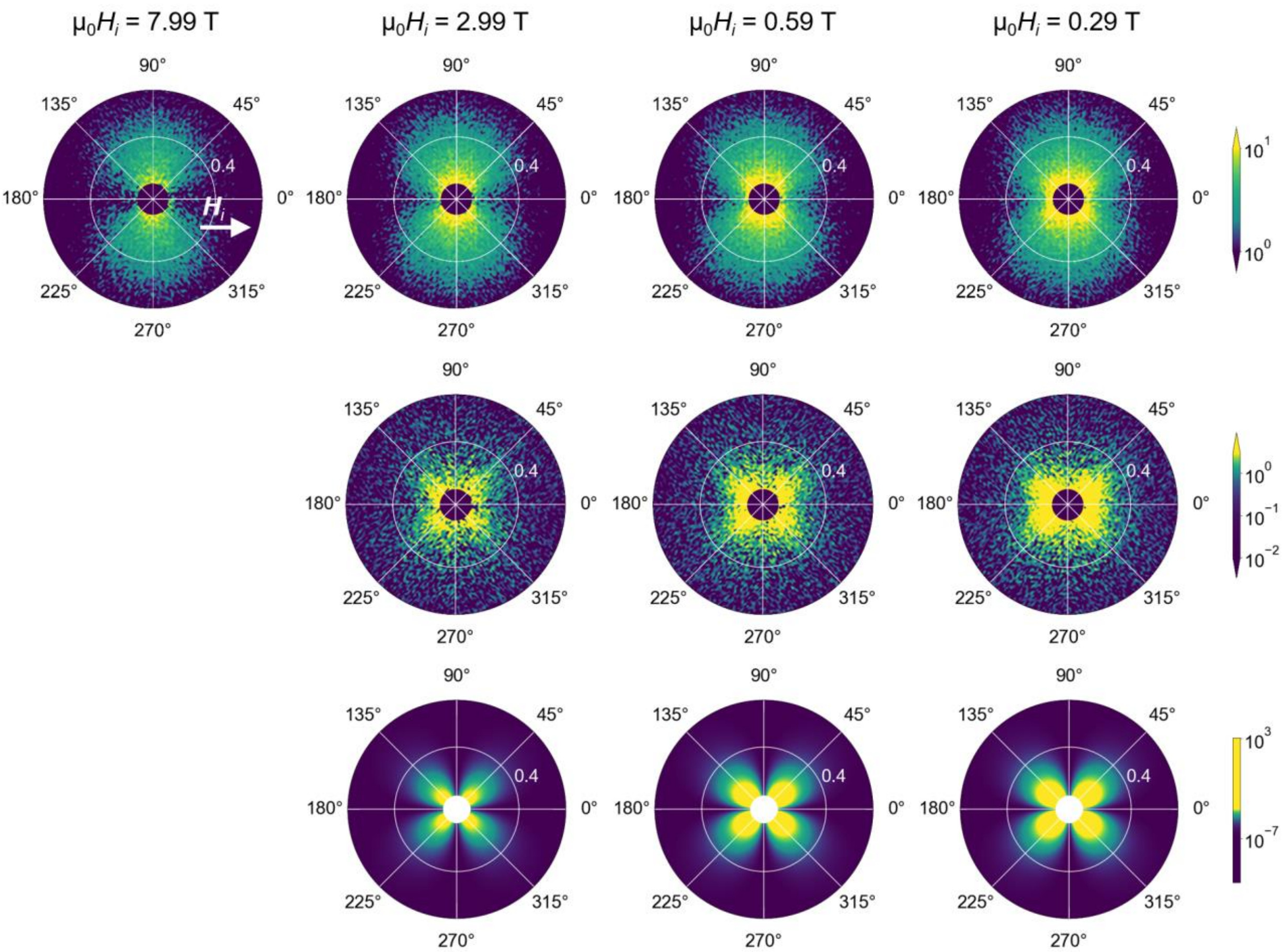

Figure 4: Experimental two-dimensional total (nuclear + magnetic) small-angle neutron scattering (SANS) cross section $d\Sigma/d\Omega$ of $(Fe_{0.7}Ni_{0.3})_{86}B_{14}$ alloy at the selected fields of 7.99, 2.99, 0.59 and 0.29 T (upper row), and the corresponding purely magnetic SANS cross section $d\Sigma_{mag}/d\Omega$ (middle row). The experimental $d\Sigma_{mag}/d\Omega$ were obtained by subtracting the $d\Sigma/d\Omega$ at the (near-) saturation field of 7.99 T from the data at the lower fields. The internal magnetic field $\boldsymbol{H}_i$ is horizontal in the plane of the detector $(\boldsymbol{H}_i \perp \boldsymbol{k}_0)$. Lower row: Computed $d\Sigma_{mag}/d\Omega$ based on the micromagnetic SANS theory [Eqs. (5)–(9)] at the same selected field values as above, and using the experimental parameters summarized in Table 1. Note that $d\Sigma/d\Omega$ and $d\Sigma_{mag}/d\Omega$ are plotted in polar coordinates with $q$ in nm$^{-1}$, $\theta$ in degree, and the intensity in cm$^{-1}$.

# FIGURE 5

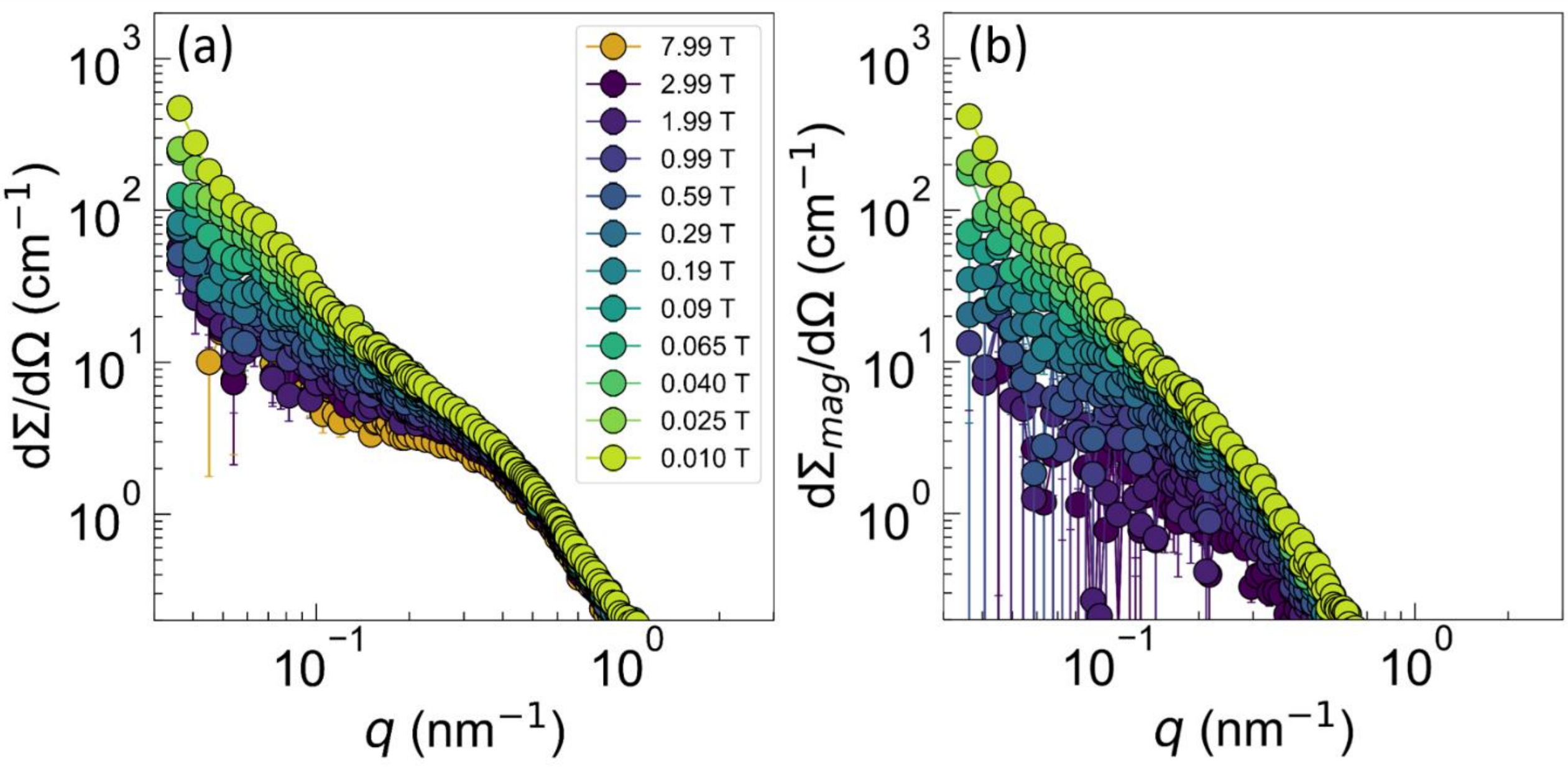

Figure 5: (a) Magnetic-field dependence of the (over $2\pi$) azimuthally-averaged total (nuclear + magnetic) SANS cross section $d\Sigma/d\Omega$ of $(Fe_{0.7}Ni_{0.3})_{86}B_{14}$ alloy. (b) The corresponding purely magnetic SANS cross section $d\Sigma_{mag}/d\Omega$ (log-log scale).

# FIGURE 6

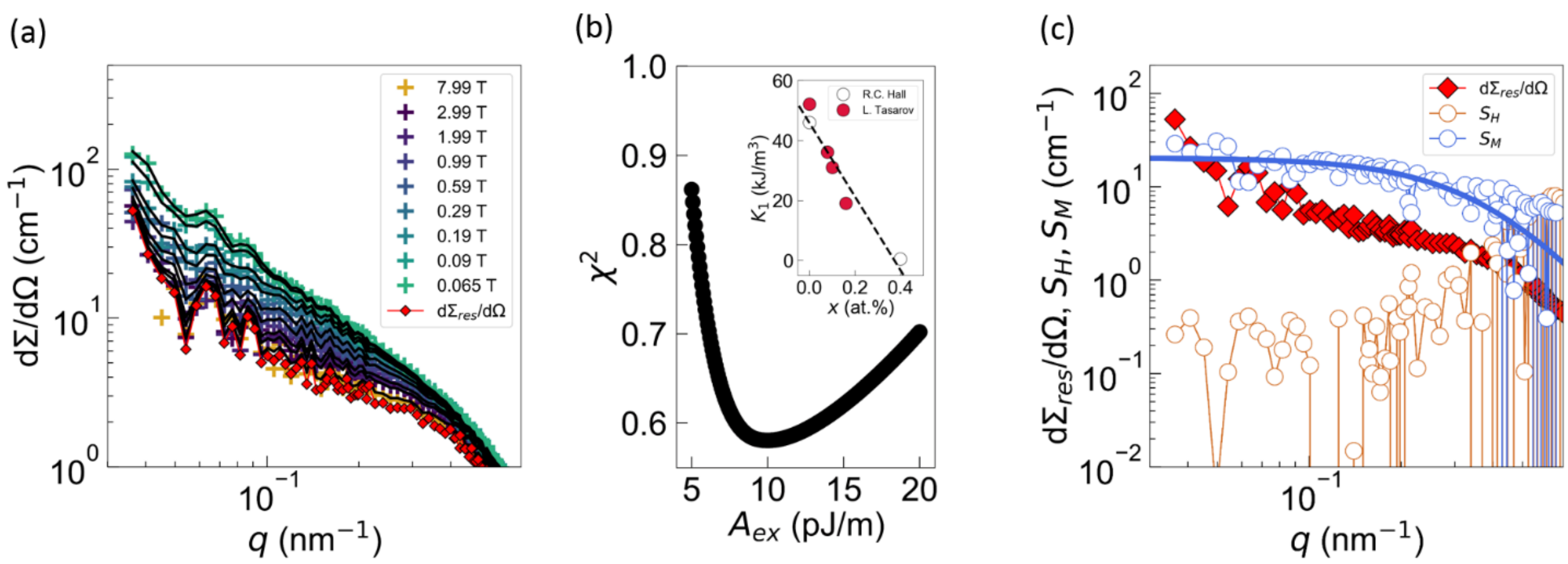

Figure 6: Results of the SANS data analysis of $(Fe_{0.7}Ni_{0.3})_{86}B_{14}$ alloy. (a) Magnetic-field dependence of the (over 2π) azimuthally-averaged total (nuclear + magnetic) SANS cross section $d\Sigma/d\Omega$ plotted in Fig. 5(a) along with the corresponding fits (black solid lines) based on the micromagnetic SANS theory [Eq. (10)]. (b) Weighted mean-squared deviation between experiment and fit, $\chi^2$, determined using Eq. (13) as a function of the exchange-stiffness constant $A_{\mathrm{ex}}$. Inset: Fe composition dependence of the magnetocrystalline anisotropy $K_1$ in $Fe_{1-x}Ni_x$ alloys (data taken from Refs. (Tarasov, 1939; Hall, 1960)). Black dashed line: Linear regression of $K_1(x)$. (c) Best-fit results for the residual scattering cross section $d\Sigma_{\mathrm{res}}/d\Omega$ (red diamonds), the scattering function $S_H$ (orange open circles), and $S_M$ (blue open circles). Blue solid line: Fit of $S_M$ assuming a Lorentzian-squared function for the $q$-dependence.

# TABLE 1

| Parameter | $(Fe_{0.7}Ni_{0.3})_{86}B_{14}$ alloy | unit |
|---|---|---|
| $a$ | ≈ 0.359 | nm |
| $D$ | 7 ± 1 | nm |
| $\mu_0 M_S$ | 1.34 ± 0.20 | T |
| $\mu_0 H_C$ | ≈ 0.0049 | mT |
| $T_c^{\mathrm{am}}$ | 720 | K |
| $A_{\mathrm{ex}}$ | 10 ± 1 | pJ/m |
| $\xi_M$ | 2.4 ± 0.2 | nm |
| $L_0$ | ≈ 50 | nm |
| $\mu_0 \langle \lvert H_p \rvert^2 \rangle$ | ≈ 0.3 | mT |
| $\mu_0 \langle \lvert M_z \rvert^2 \rangle$ | ≈ 24 | mT |

Table 1: Summary of the structural and magnetic parameters for $(Fe_{0.7}Ni_{0.3})_{86}B_{14}$ alloy (HiB-NANOPERM-type soft magnetic nanocrystalline material) determined by wide-angle x-ray diffraction, magnetometry, differential thermal analysis, thermo-magneto-gravimetric analysis, and small-angle neutron scattering.

# Footnote 1

Estimated by assuming a linear regression of $K_1$ in $Fe_{1-x}Ni_x$ alloys for a Fe composition $x$ between 0 and 0.4 at. % (see inset in Fig. 6(b), data taken from Refs. (Tarasov, 1939; Hall, 1960).